\documentstyle[12pt,psfig]{article}
\begin{document}

\centerline{\Large \bf On the accuracy of GAIA radial velocities}

\vskip 0.5 cm
\centerline{\bf U. Munari, P. Agnolin and   
L. Tomasella\footnote{send off-print requests to U.Munari, 
{\tt munari@pd.astro.it}}}

\vskip 0.5 cm
\centerline{Osservatorio Astronomico di Padova, Sede di Asiago, 
I-36012 Asiago (VI), Italy}

\vskip 1.0 cm \noindent
{\sl {\rm \bf Abstract.}

We have obtained 782 real spectra and used them as inputs for 6700 automatic
cross-correlation runs to investigate GAIA potential in terms of radial
velocity accuracy. We have explored the dispersions 0.25, 0.5, 1 and 2
\AA/pix over the 8490--8740 \AA\ GAIA range. We have investigated late-F to
early-M stars (constituting the vast majority of GAIA targets), slowly
rotating ($<\, V_{rot}\, \sin\, i\,>$ = 4 km~sec$^{-1}$), of solar
metallicity ($<$\,[Fe/H]\,$>$ = $-$0.07) and not binary. The results are
accurately described by the simple law: {\tt $\lg\, \sigma\, =\,
0.6\times(\lg\frac{S}{N})^2\, -\, 2.4\times\lg\frac{S}{N}\, +\, 1.75\times
\lg D\, +\, 3$}, where $\sigma$ is the cross-correlation standard error (in
km sec$^{-1}$) and $D$ is the spectral dispersion (in \AA/pix).  The
spectral dispersion has turned out to be the dominant factor, with S/N being 
less important and the spectral mis-match being a weak player at the 
lowest S/N. Our results show that mission-averaged radial velocities of faint 
GAIA targets (V$\sim$15 mag) can match the $\sim$0.5 km~sec$^{-1}$ 
accuracy of tangential motions, provided the observations are performed 
at a dispersion not less than 0.5
\AA/pix.}

\section{Introduction}

GAIA is the coming ESA Cornerstone 6 mission. It is designed to obtain
extremely precise astrometry (in the {\sl micro-}arcsec regime), multi-band
photometry and medium/high resolution spectroscopy for a large sample of
stars.  The goals as depicted in the mission {\sl Concept and Technology
Study Report} (ESA SP--2000--4, hereafter {\sl CTSR}) call for astrometry
and broad band photometry to be collected for all stars down to $V \sim$20
mag over the entire sky ($\sim 1$ 10$^9$ stars), with brighter magnitude
limits for spectroscopy and intermediate band photometry. Each target star
should be measured about a hundred times during the five year mission
life-time, in a fashion similar to the highly successful {\sl Hipparcos}
operation mode. The astrophysical guidelines of the GAIA mission are
discussed by Gilmore et al. (1998, 2000) and Perryman et al. (2001), an
overview of the GAIA payload and spacecraft is presented by M\'erat et al.
(1999), while the goals of the GAIA spectroscopy are
discussed by Munari (1999, 2001).

The principal aim of GAIA spectroscopy will be to provide the 6$^{th}$
component of the phase-space, the radial velocity. The obvious goal of GAIA
radial velocities is to parallel the precision of tangential motions. The
latter is a combination of the precision of parallaxes and proper motions.
From stellar population models of the Galaxy, {\sl CTSR} estimates an
average accuracy of 0.5 km sec$^{-1}$ for the tangential motions of $V\sim
15$ mag stars. The latter is close to the magnitude limit of GAIA
spectroscopy, the exact one depending on the final optical design, on-board
data processing strategies, detection threshold, telemetry constraints and
so forth. To appropriately complement the keen astrometric vision of the
Galaxy, GAIA radial velocities therefore need to be accurate at their faint
magnitude limit.

GAIA will record spectra covering the 8490--8740 \AA\ region (cf. Munari
1999), at 0.75 \AA/pix dispersion as currently baselined in the {\sl CTSR}.
A $\bigtriangleup \lambda \sim$ 250 \AA\ range is mainly imposed by optical
constraints on the maximum $\lambda-$interval well focusable over the large
field ($2^\circ \times 1^\circ$) of the GAIA spectroscopic focal plane,
where CCDs operated in TDI mode will record the spectra as they transit over
the field. The GAIA 8490--8740 \AA\ interval is dominated by the Ca~II
triplet, which is among the strongest spectral lines at any optical
wavelength in F-G-K-M stars. The latter will be the dominating types among
GAIA spectroscopic targets (from Galaxy models and Hipparcos/Tycho data, the
average color of field stars at $V=10$ mag corresponds to a G0 spectral type,
and to K0 at $V=15$ mag).

\begin{figure*}
\centerline{\psfig{file=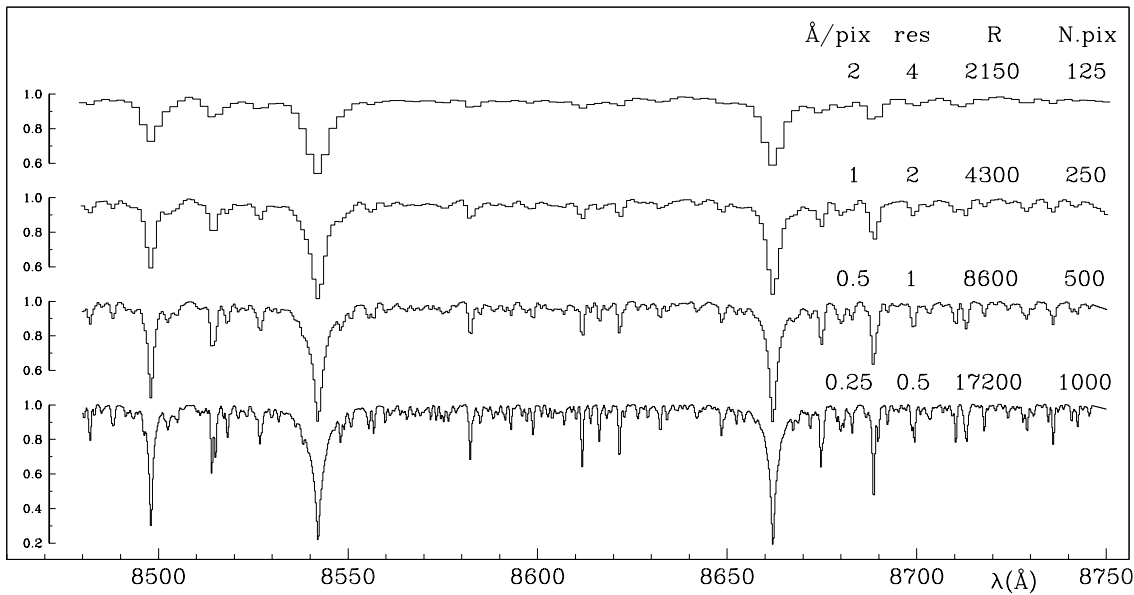,angle=270,width=15cm}}
\caption{The figure illustrates the effect of degrading the dispersion of
GAIA spectra (from one spectrum to the next the dispersion changes by a
factor of 2). The 0.75 \AA/pix currently baselined for GAIA falls at the
center of the explored 8$\times$ range. To focus upon the features carrying
RV informations and to avoid disturbance by the necessarily limited S/N of
real spectra, synthetic spectra from Munari and Castelli (2000) have been
used to produce this figure. The spectra refer to a K0~III star with T=4750~K,
$\lg$ g=3.0, [Z/Z$_\odot$]=0.0 
and $V_{rot} \sin\, i=$5 km
sec$^{-1}$, i.e. the average program star in Table~1. Over each spectrum the
dispersion (in \AA/pix), the resolution (in \AA), the resolving power $R$
and the spectrum length (in pixels) are given.}
\end{figure*}

The aim of the present paper is to evaluate the {\sl potential} precision of
GAIA radial velocities as function of spectral resolution and
signal-to-noise ratio. How much of this {\sl potential} precision will be
effectively exploited by GAIA will depend upon the control over the
wavelength scale of the recorded spectra, which is open to refinements (the
{\it modus operandi} of GAIA spectroscopy being similar to an
``objective-grism" approach, where for obvious reasons calibration lamps
and/or telluric absorptions cannot be used as in classical ground-based
spectra). In this paper we use real data (obtained with real CCDs at
real spectrographs+telescopes) and automatic cross-correlation measurement
of radial velocities in the attempt to better account -- compared to
simulations -- for all sources of noise, manifest as well as hidden,
affecting GAIA observations and data reduction.

This study follows and confirms preliminary investigations we performed and
circulated in 1997--98 as internal documents of the ESA Photometric and
Spectroscopic Working Group for GAIA (UM-PWG-005, UM-PWG-006).

\section{Data}

\subsection{Selection of program stars}

The late-F to early-M program stars have been selected -- according to
visibility at the time of the observing runs -- among IAU standard RV stars
(as listed in the Astronomical Almanac). Three additional bright stars (with
no record of binarity, radial velocity variability or spectral
peculiarities) were selected to complement the observations at 0.25 \AA/pix.

The program stars are listed in Table~1, together with their RVs (from
Astronomical Almanac, 2002 edition), rotational velocity (from Glebocki et
al. 2000) and metallicity (from Cayrel de Strobel et al. 1997). Their median
rotational velocity is $V_{\rm rot}\,\sin\,i$ = 4 km~sec$^{-1}$, which is
close to the median value of field stars as shown in Table~2.

\subsection{Explored dispersions and resolutions}

We have explored four dispersions: 0.25, 0.5, 1 and 2 \AA/pix. They bracket
the 0.75 \AA/pix currently baselined for GAIA.

Figure~1 is a visualization of these dispersions upon the same K0~III
synthetic spectrum (with $V_{\rm rot}\,\sin\,i$ = 4 km~sec$^{-1}$ and
[Z/Z$_\odot$] = 0.0, from Munari and Castelli 2000). Throughout this paper the
spectral resolution (taken as the FWHM of the PSF) is kept constant at 2.0
pixels. Therefore the resolutions explored in this paper are 0.5, 1, 2 and 4
\AA\ corresponding to resolving powers $R$=17\,200, 8\,600, 4\,300 and
2\,150.

\begin{table}
\centerline{\psfig{file=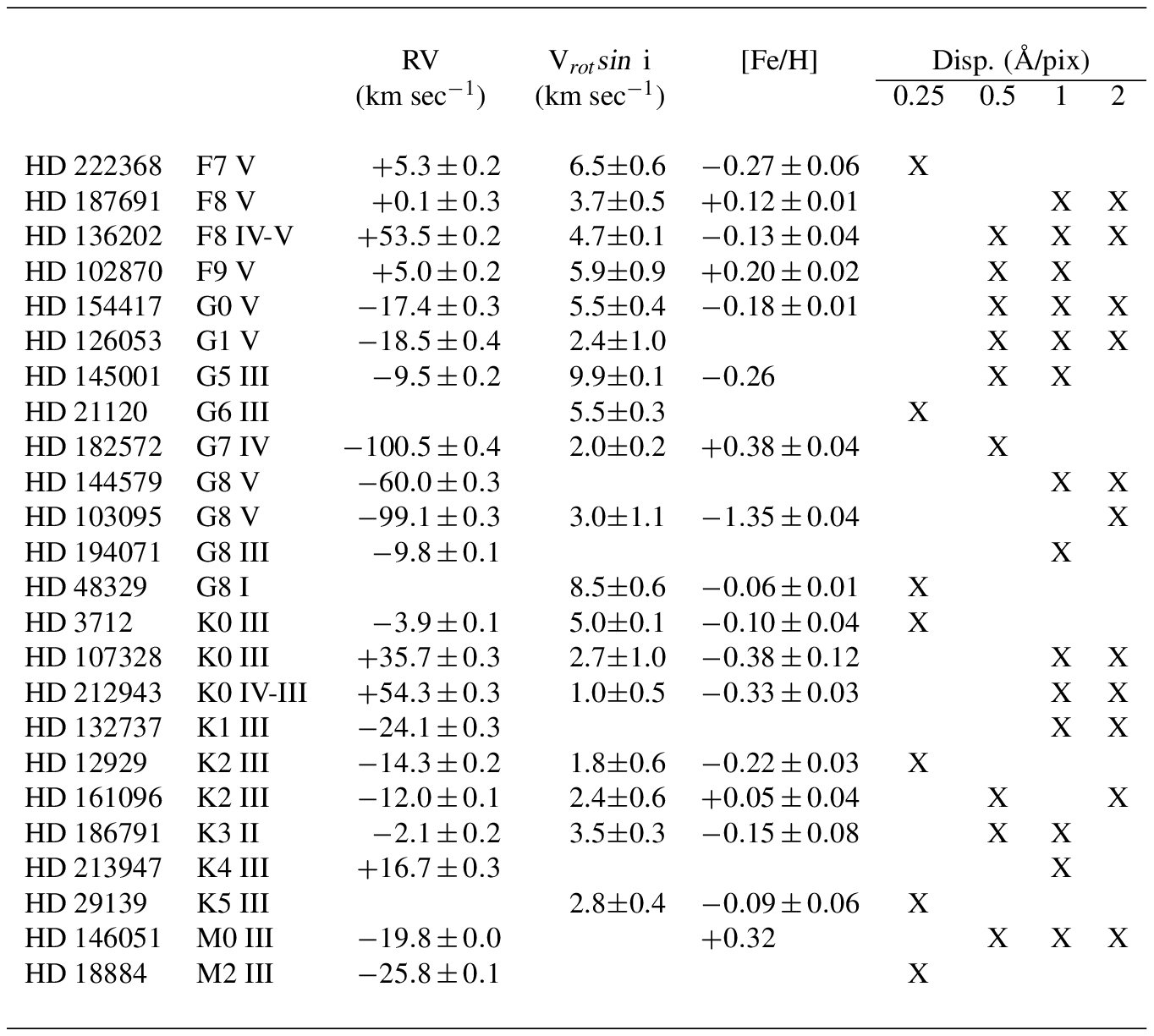}}
\caption[]{Program stars. The radial velocities are from {\sl Astronomical
Almanac 2002}, rotational velocities from Glebocki et al. (2000) and
metallicities from Cayrel de Strobel et al. (1997). The last four columns
indicate the dispersions at which the program stars have been observed.}
\end{table}
\begin{table}
\centerline{\psfig{file=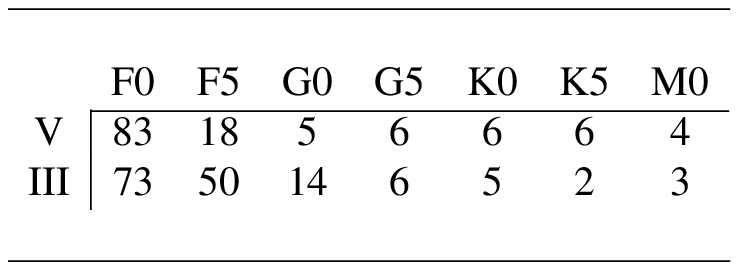}}
\caption[]{Median $V_{\rm rot}\sin\,i$ (in km~sec$^{-1}$) are computed for
selected spectral classes from the data in the Glebocki et al. (2000)
catalogue of rotational velocities for about 12\,000 field stars. Stars of
luminosity class I stars are generally undersampled in the catalogue and
thus not considered in this table.}
\end{table}

\subsection{Strategy}

The mean uncertainty upon the RVs of the IAU standard stars in Table~1 is
0.23 km sec$^{-1}$ which corresponds to a mean uncertainty of 0.33 km
sec$^{-1}$ in the RV difference between two program stars. This is
comparable to the expected error of the cross-correlation at the highest
dispersion and S/N here investigated. For this and other reasons, to
estimate the error of the cross-correlation we decided to proceed in a way
independent from the exact knowledge of the individual RVs. It only requires
the RV constancy of the given program star during the 0.5-1.5 hours of
observation, which is certainly the case for IAU standard RV stars.

For each dispersion and program star we obtained three deep spectra that were
combined into a single high S/N {\sl template} (S/N$\geq$220). We then
proceeded for each program star to obtain five spectra at S/N$\sim$110,
five spectra at S/N$\sim$35 and other five spectra at S/N$\sim$10 by
properly adjusting the exposure time. The final extracted spectra had an
average S/N per pixel of the continuum of 110, 33 and 12, respectively.

For each program star we then cross-correlated the five spectra at
S/N$\sim$110 against the high S/N template and derived a standard deviation
of the resulting radial velocities. The same was done for the five
S/N$\sim$33 spectra and for the five at S/N$\sim$12. To increase the
statistics and to account for possibly large mis-matches between objects and
templates in the GAIA automatic cross-correlations, the 5+5+5 {\sl object}
spectra of each 

\clearpage

\noindent
given star were cross-correlated against the {\sl template}
spectra of all program stars (at the given dispersion) and the average
standard deviation is given in Table~3.

\subsection{Observations}

\begin{table*}
\centerline{\psfig{file=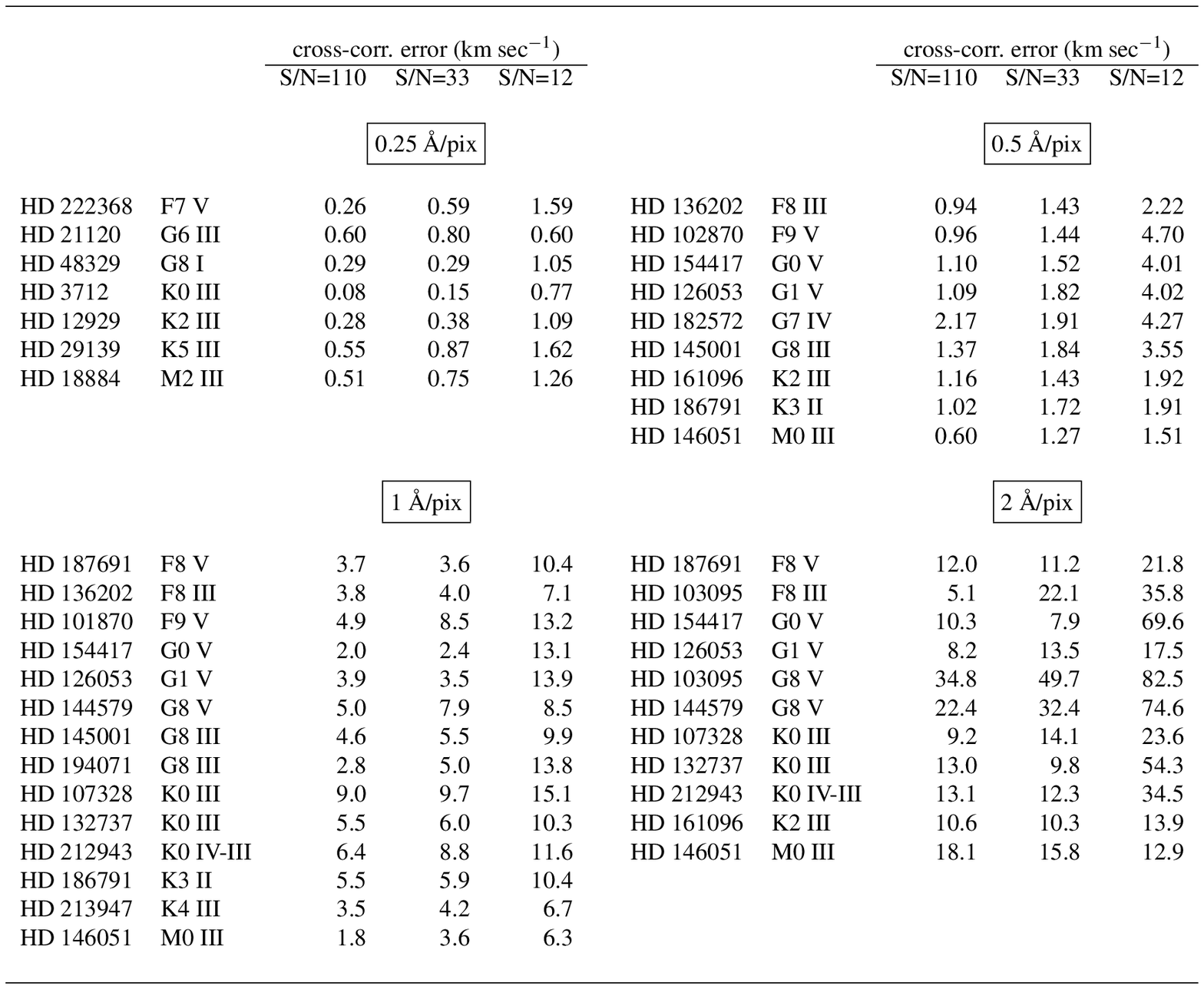,width=15cm}}
\caption[]{Errors of radial velocities obtained via cross-correlation (in
km~sec$^{-1}$). The values reported are the mean of the standard deviations
of the cross-correlation results when the spectra of the given program star
are cross-correlated against the templates of {\sl all} program stars at the
given dispersion.}
\end{table*}

The observations have been obtained during four nights in 2000 (Jun 15, 21
and 22) and 2001 (Jan 14). The spectra at a given dispersion have been all
secured during the same night under identical and very stable instrumental
conditions. On a few more nights the observing program could not be
completed because of intervening clouds, and data obtained under such
circumstances have not been further used in this paper. To complete the
observations on a given star at a given dispersion, about one hour was
generally enough.

The 0.25 and 0.5 \AA/pix pointed observations have been obtained with the
Echelle+CCD spectrograph attached to the 1.82 m telescope operated in Asiago
by Osservatorio Astronomico di Padova (Italy). An OG455 filter was used to
eliminate the cross-disperser second order. The 8490-8740 \AA\ range is
covered by 1000 pixels (500 for the 0.5 \AA/pix dispersion) and it is fully
contained within one Echelle order. For both dispersions the spectrograph
slit width was adjusted so that FWHM(PSF) = 2.0 pixels.

The 1 and 2 \AA/pix pointed observations were secured with the B\&C+CCD
spectrograph attached to the 1.22 m telescope operated at Asiago by the
Dept. of Astronomy of the University of Padova (Italy). A 1200 and a 600
ln/mm gratings were used together with an RG1 filter to eliminate the grating
second order. Again, for both dispersions the spectrograph slit width was
adjusted so that FWHM(PSF) = 2.0 pixels.

\subsection{Cross-correlations}

The cross-correlations were performed via the {\tt fxcor} task in IRAF,
operated in automatic batch mode (plus some custom codes in C++). A total of
6705 cross-correlation runs were performed on the 782 input spectra.

We used pixel-based extracted spectra, without wavelength calibration to
avoid to introduce spurious effects and to better simulate the GAIA data.
Observations obtained with a Cassegrain telescope suffer from thermal
changes and mechanical flexures, which will be absent in GAIA. Accurate
cross-correlation of the rich telluric absorption forest (cf. Munari 1999)
flanking on both side the GAIA wavelength region has been used to compensate
for the the spectrograph's focal plane drift over the CCD. The minimal
thermal changes in the dome during the observations produced unnoticeable
focal plane drifts, while the mechanical flexures could always been
compensated for by cross-correlation between the telluric lines in the {\sl
template} and {\sl object} spectra. The mechanical flexures matched their
mathematical modeling by Munari and Lattanzi (1992).

\section{Results}

To further filter out the noise due to limited number statistics and to
better characterize the role of dispersion and S/N, the star-by-star data of
Table~3 are averaged in Table~4. The latter can be simply read as the
standard error -- at a given dispersion and S/N -- of the cross-correlation
between an object and a template randomly chosen among late-F/early-M
spectral types at any luminosity class (with average $V_{\rm rot}\,\sin\,i$
= 4 km~sec$^{-1}$ and [Z/Z$_\odot$] = 0.0). Table~5 lists for each
dispersion the magnitudes of the stars observed by GAIA that provide spectra
of the given S/N per single passage over the field of view.

\subsection{The role of spectral dispersion and S/N}

The data in Table~4 are well fitted by the simple law:
\begin{equation}
\lg\ \sigma\ =  \  0.6\times (\lg \frac{S}{N})^2\ 
                 -\  2.4\times \lg \frac{S}{N}\ 
                 +\ 1.75\times \lg D\ +\ 3
\end{equation}
where $\sigma$ is the standard error in km sec$^{-1}$,  
$\frac{S}{N}$ is obviously the signal-to-noise ratio (per pixel on the
stellar continuum) and $D$ is the spectral dispersion (in \AA/pix).

The effect of S/N (at least over the 12 -- 110 range here explored) is
pretty similar at all investigated dispersions: going from S/N=12 to S/N=33
doubles the accuracy of RVs, while an increase from S/N=33 to S/N=110
increases the precision of RVs by only 35\%.

It is therefore clear from Table~4 and Figure~2 that the dispersion is the
principal factor governing the potential accuracy of GAIA radial velocities,
with S/N playing a less important role: pushing the exposure time so long to
achieve S/N$\sim$110 does not provide more accurate results than obtainable
with S/N$\sim$12 spectra at twice better resolution.

\begin{table}
\centerline{\psfig{file=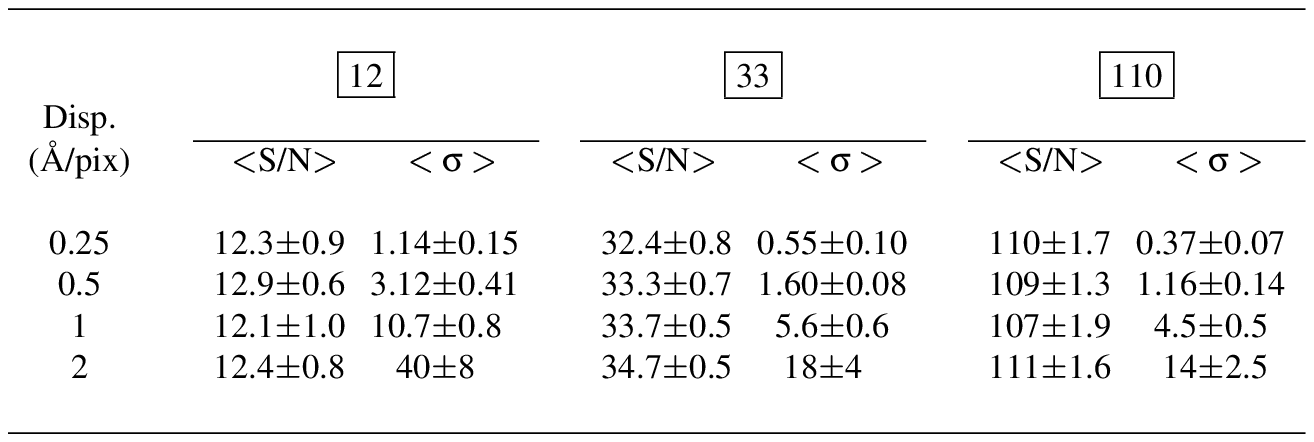}}
\caption[]{The accuracy of radial velocities obtained via cross-correlation
for late-F to early-M stars as function of S/N and spectral dispersion. The
reported values are grand-averages over the results of Table~3 and the
uncertainties are formal errors of the mean. {\sl 12}, {\sl 33} and
{\sl 110} refer to the average S/N of each group.}
\end{table}
\begin{figure}
\centerline{\psfig{file=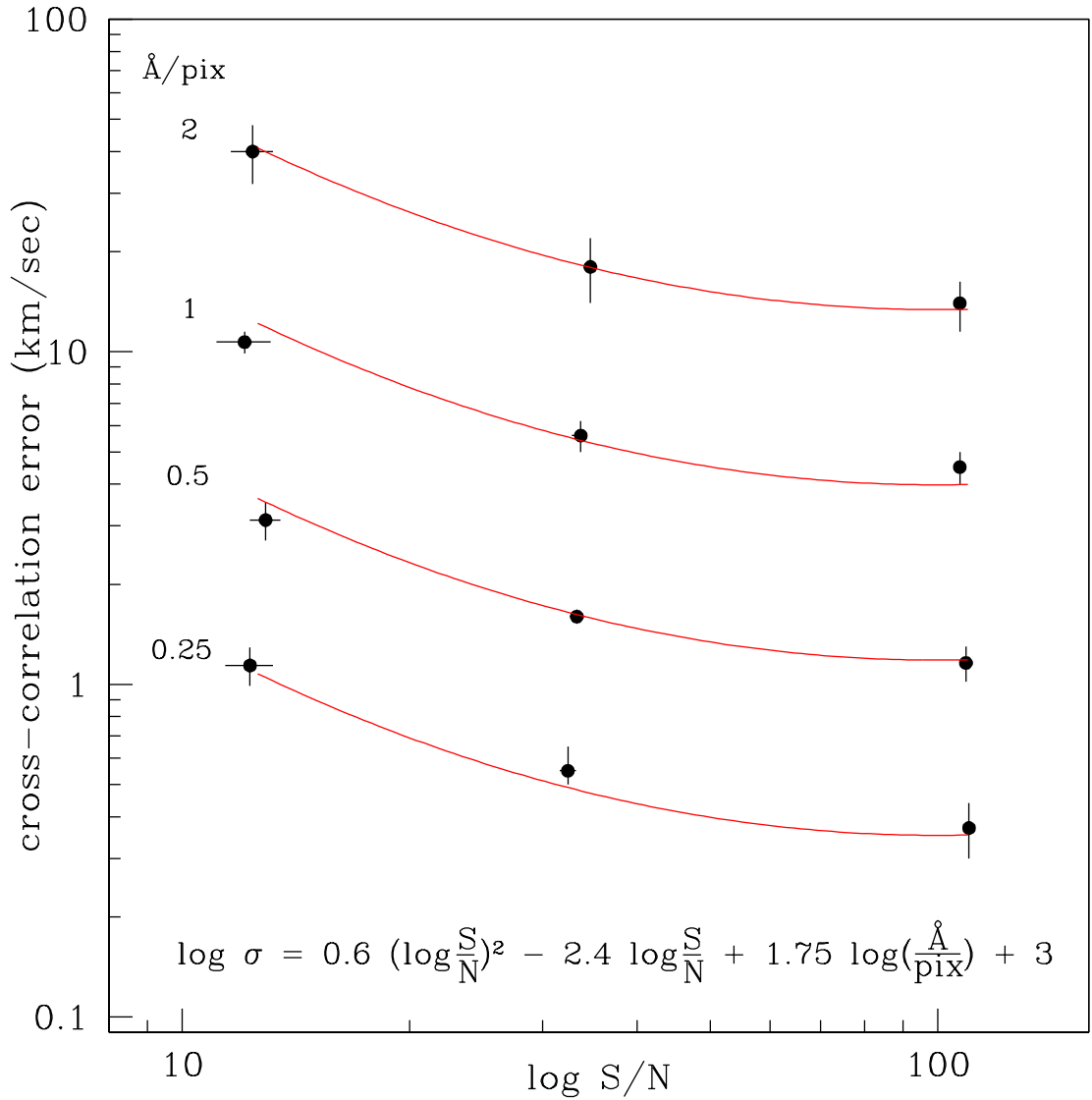,width=15cm}}
\caption[]{The results of Table~4 in graphical form. The results for
each individual dispersion show the same dependence upon S/N. This
is emphasized by the fitting curves obtained from the expression for 
$\lg \sigma$ by inserting the appropriate value of the dispersion
(in \AA/pix).}
\end{figure}

\subsection{The effect of mis-match}

Equation (1) is affected by the mis-match between {\sl template} and
{\sl object} spectra. To account for it too would have required an
unrealistic observational effort (about 30$\times$ more spectra, in the
neighborhood of 25\,000 total). Therefore, the systematic exploration of
the (minor) role played by mis-match in spectral type, metallicity and
rotational velocity can only be performed via cross-correlation of huge
databases of {\sl synthetic} spectra (Zwitter 2001, in preparation). A few
preliminary considerations are however in order, separately for low and high
S/N spectra.

For S/N=12 and S/N=33 spectra, we can roughly and preliminary estimate it
only for K0/K1 program stars and for 1 and 2 \AA/pix dispersions, for which
at least three such stars were observed. A check on the original data (not
reported here for their excessive length) show that cross-correlating 1 or 2
\AA/pix spectra of K0/1~IV/III program stars with templates of the same
spectral types produces $\sim$15\% better results than cross-correlation
with templates of largely different spectral types chosen between the
late-F/early-M boundaries. Such a result is readily appreciable when the 1
and 2 \AA/pix spectra in Figure~1 are considered: at these low 

\begin{table} 
\centerline{\psfig{file=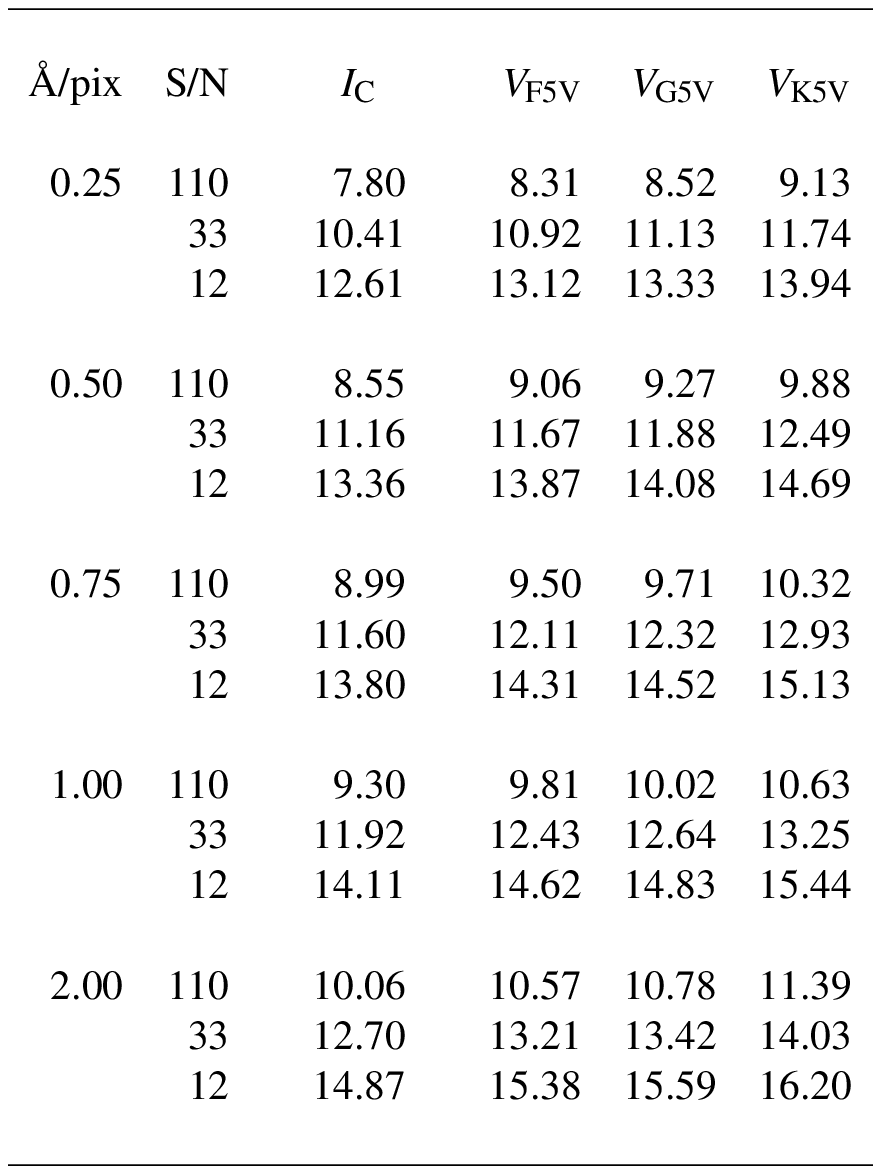}} 
\caption[]{The table provides the magnitudes of the stars observed by GAIA
that correspond to the explored S/N (per pixel) at the given dispersion per
single passage over the field of view. The magnitudes are computed for the
Cousins' $I$ band, which covers the wavelength range of GAIA spectra. The
corresponding $V$ magnitudes are listed for F5~V ($V-I_{\rm C}$=+0.51), G5~V
($V-I_{\rm C}$=+0.72) and K5~V ($V-I_{\rm C}$=+1.33) unreddened stars. The
following parameters have been adopted in the computations: mirror size =
75$\times$70 cm; overall throughput = 35\%; crossing time = 60.8 sec; $I_{\rm
C}^{mag=0.0}$ = 1.196$\times 10^{-9}$ erg cm$^{-2}$ sec$^{-1}$ \AA$^{-1}$ = 519
photons cm$^{-2}$ sec$^{-1}$ \AA$^{-1}$; R.O.N. = 3 $e^{-1}$; dark = 0.01 
$e^{-1}$ sec$^{-1}$; sky background $I_{\rm C}$=21.5 mag arcsec$^{-2}$.}
\end{table}

\noindent
dispersions only the three Ca~II and a few Fe~I lines carry RV information and
the same lines are always and invariably the dominating features in the
spectra of the late-F to early-M stars (cf. the GAIA spectral atlases of
Munari and Tomasella 1999, Munari and Castelli 2000 and Castelli and Munari
2001).

For S/N=110 spectra, the mis-match is expected to be proportionally more
important.  Whatever high the S/N might be, unavoidable differences remain
in {\sl template} and {\sl object} spectra simply because they are
intrinsically different and this places a physical limit to the accuracy the
cross-correlation can achieve.

In conclusion, in absence of mis-match it could be expected the fitting
lines in Figure~2 to be less curved and more straight at the higher S/N,
maintaining however the same spacing among them and the same slope at the
lower S/N. Put in other words, a proper account for mis-match would have
only minor effects at low S/N while it is expected to improve more
significantly the accuracy achievable with high S/N data.  Therefore, to
exploit the best intrinsic accuracy of the highest signal-to-noise GAIA
spectra it will be necessary to accurately determine and use the best
possible template for each individual spectrum (that will however concern a
minimal part of the GAIA spectra for which an {\sl ad hoc} reduction
pipeline can be envisaged if necessary).

\subsection{Instrumental stability}

A careful control over the instrumental conditions and stability was
maintained during the acquisition of the 782 input spectra used for this
paper. As a side test, we obtained a further set of 0.25~\AA/pix spectra
under non-stable conditions (typically one object per night between June 17
and June 25 2000, with grating and slit moved back and forth and
spectrograph refocused during day-time, and with appreciable thermal cycles
$\bigtriangleup T \sim$4--8 $^\circ$C from day to night; several colleagues
kindly offered a short part of their observing time for the purpose). The
data were then extracted and the spectra cross-correlated in exactly the
same manner as described for the main data. The result of these additional
0.25 \AA/pix observations, aimed to simulate poor environmental and
instrumental control over GAIA observations, are: $<\sigma>$=2.95
km~sec$^{-1}$ at $<$S/N$>$=14, $<\sigma>$=1.53 at $<$S/N$>$=35 and
$<\sigma>$=1.03 at $<$S/N$>$=106. They are only marginally better than the
values reported in Table~4 for the 0.5 \AA/pix dispersion. A poor
thermal/mechanical stability has therefore vanished the gain expected from a
twice higher dispersion. This (obvious) result stresses the advantages of
the unique and very favorable GAIA environment, characterized by extreme
mechanical and thermal stability.

\section{Discussion}

Our investigation has shown that mission-averaged radial velocities of faint
GAIA targets (V$\sim$15 mag) can match the $\sim$0.5 km~sec$^{-1}$ accuracy
of tangential motions, provided the observations are performed at a
dispersion not less than 0.5 \AA/pix. From Table~4 and Figure~2 a S/N=10
spectrum at 0.5 \AA/pix dispersion has a standard error of $\sim$5
km~sec$^{-1}$, and a hundred such spectra are necessary to lower the error
on the mean velocity to 0.5 km~sec$^{-1}$. This is valid for {\sl single}
stars of the dominating GAIA target population, i.e. slow rotating late-F to
early-M stars. Binary and/or hotter and/or faster rotating and/or pulsating
stars will require a higher accuracy for individual radial velocities (and
thus presumably a higher dispersion) to match the $\sim$0.5 km~sec$^{-1}$
precision of GAIA tangential motions at V$\sim$15 mag.

The results of the present observational study agrees with the findings of
numerical simulations cited in the {\sl CTSR} that call for mission-averaged
RVs of faint GAIA cool targets to be accurate to $\sim$5 km~sec$^{-1}$. This
agrees with the results in Table~4 and Figure~2 for the 0.75 \AA/pix
dispersion baselined for GAIA when proper allowance is made for possible
binary nature and/or pulsational activity of the target stars.

The way GAIA will control the zero-point of the wavelength scale of its
spectra reinforces the quest for a high spectral dispersion. The
determination of the zero-point will be provided by ($a$) centering of the
zero-order image on the spectrograph focal plane, and/or ($b$) accurate
knowledge of the focal plane geometry (as mapped by standard RV stars
observed during the mission) + astrometric position of the target stars The
accuracy of both methods can be expressed as a centering error in pixels,
and it is obvious that the higher the spectral dispersion the less
km~sec$^{-1}$ correspond to a 1-pixel shift.

We have shown how the spectral dispersion is the key factor in governing the
accuracy of radial velocities. To increase the S/N per pixel on the stellar
continuum from S/N$\sim$12 to S/N$\sim$110 requires $\sim 80\times$ more
photons. However, a S/N$\sim$110 spectrum cannot provide a radial velocity
significantly more accurate than a S/N$\sim$12 spectrum obtained at twice
higher resolution, which costs only $2\times$ more photons.

It is also worth to remind that the higher the dispersion, the wider the
usage and interest of the spectra. Spectra of high enough dispersion allow
-- for instance -- rotational velocities and chemical abundances to be
measured with confidence. Lowering the dispersion in favor of a higher S/N
per pixel (GAIA exposure time is the same for all stars and set by the 
spacecraft axial
rotation period and angular extent of the imaged field of view) does not
seem a viable alternative. In fact, when the rotation broadening become
sub-pixel it cannot be reliably measured (15 km~sec$^{-1}$ corresponds to
0.43 \AA\ at GAIA wavelengths, and the majority of GAIA targets will be
rotating slower than 15 km~sec$^{-1}$, cf. Table~2). Similarly, a low
dispersion means severe line blending and the impossibility of
element-by-element chemical analysis. On the other hand the higher the
dispersion, the more complicate the spectrograph realization and heavier the
demands on down-to-Earth telemetry budget. It is easy to anticipate that a
lot of work will still be necessary during the final design phase for GAIA
to find the best compromise between science demands and technical
challenges.

Hot stars (O, B and A types) have not been considered in this paper because
they will account for a small fraction of GAIA targets. Preliminary results
that we have circulated in 1997 and 1998 among the ESA Photometric and
Spectroscopic Working Group for GAIA show that the accuracy of radial
velocity rapidly degrades moving toward hotter spectral types (Paschen lines
are a weaker spectral feature than Ca~II triplet lines). These
preliminary results need however refinements and to this aim we have already
started the acquisition at the telescope of a large sample of suitable
spectra of O, B and A stars to be used in a coming investigation paralleling
the present one. Finer analysis of the role of rotation, spectral mis-match
and metallicity (the higher the metallicity, the stronger the absorption
lines and therefore the stronger the radial velocity signature in the
cross-correlation) are also in order and will be considered elsewhere (Zwitter 2001,
in preparation).

Finally, it is also worth to remind that these results are relevant not only
within the GAIA context but also to ground-based observers because the
absence of telluric absorptions and proximity to the wavelengths of peak
emission make the explored 8490--8740 \AA\ interval an interesting option
for studies of cool stars with conventional telescopes.

\vskip 1.0 cm
\noindent
{\sl Acknowledgements.}
We would like to thank R.Barbon, T.Zwitter and T.Tomov for useful discussions 
and comments on an early draft of the paper.

\vskip 1.0 cm \noindent 
{\Large \bf References}
\vskip 1.0 cm
\noindent
Castelli F., Munari U. 2001, A\&A 366, 1003\\
Cayrel de Strobel G., Soubiran C., Friel E.D., Ralite N., Francois P. 1997, 

A\&AS 124, 299\\
Gilmore G., Perryman M., Lindegren L., Favata F., Hoeg E., Lattanzi M., Luri X., 

Mignard F., Roeser S., de Zeeuw P.T., 1998, Proc SPIE Conference 3350, p. 541\\
Gilmore, G. F., de Boer, K. S., Favata, F., Hoeg, E., Lattanzi, M. G., 

Lindegren, L., Luri, X., Mignard, F., Perryman, M. A. C., de Zeeuw, P.T. 2000

Proc SPIE Conference 4013, p. 453\\
Glebocki R., Gnacinski P., Stawikowski A. 2000, Acta Astron. 50, 509\\
M\'erat P., Safa F., Camus J.P., Pace O., Perryman M.A.C. 1999, in Proceedings

of the ESA Leiden Workshop on GAIA, 23-27 Nov 1998, Baltic Astronomy, 8, 1\\
Munari U. 1999, in Proceedings of the ESA Leiden Workshop

on GAIA, 23-27 Nov 1998, Baltic Astronomy, 8, 73\\
Munari U., 2001, in Proceedings of the Les Houches School "GAIA: the Galaxy Census"

ed.s C.Turon, O.Bienayme, in press\\
Munari U., Castelli F. 2000, A\&AS 141, 141\\
Munari U., Lattanzi M.G.  1992, PASP 104, 121\\
Munari U., Tomasella L. 1999, A\&AS 137, 521\\
Perryman, M. A. C., de Boer, K. S., Gilmore, G., Hoeg, E., Lattanzi, M. G., 

Lindegren, L., Luri, X., Mignard, F., Pace, O., de Zeeuw, P. T.  2001, A\&A 369, 339
\end{document}